\documentclass[prl,twocolumn,floatfix,showpacs]{revtex4}
\usepackage{psfig}
\begin{document}
\title{Inelastic X-ray scattering in correlated (Mott) insulators.}
\author{T. P. Devereaux, G. E. D. McCormack}
\affiliation{Department of Physics, University of Waterloo, Waterloo, ON, 
Canada}
\author{J. K. Freericks}
\affiliation{Department of Physics, Georgetown University, Washington, DC, USA}
\date{\today}
\begin{abstract}
We calculate the inelastic light scattering from X-rays, which
allows the photon to transfer both energy and momentum to the
strongly correlated
charge excitations. We find that the charge transfer
peak and the low energy peak both broaden and disperse through the
Brillouin zone similar to what is seen in experiments in materials
like Ca$_2$CuO$_2$Cl$_2$.
\end{abstract} \pacs{78.70.Ck, 72.80.Sk, 78.66.Nk, 71.30.+h, 71.27.+a} 
\maketitle

The dynamics of electrons in strongly correlated systems is far
from well-understood. In a Mott insulator, correlations split a
single band into a lower and an upper Hubbard band separated by a
Mott gap. Many experimental probes have focused attention on
exploring the detailed nature of the lower Hubbard band from which
electrons may be excited using angle-resolved photoemission
(ARPES) for example, but the structure and symmetry of the upper
Hubbard band and the relaxational dynamics of electrons populated
into it remains largely unexplored.

Inelastic X-ray scattering\cite{review-xray} 
has attempted to address this issue on
a number of correlated insulators such as La$_{2}$CuO$_{4}$ and
Sr$_{2}$CuO$_{2}$Cl$_{2}$ \cite{abbamonte},
Ca$_{2}$CuO$_{2}$Cl$_{2}$\cite{Hasan2000},
NaV$_{2}$O$_{5}$\cite{Na}, Nd$_{2}$CuO$_{4}$\cite{Nd}, and 1-D
insulators Sr$_{2}$CuO$_{3}$ and
Sr$_{2}$CuO$_{2}$\cite{Hasan2002}. The measured signal is
resonantly enhanced by tuning the incident photon energy to lie
near the Cu $K$ or V $L_{3}$ edge. The measurements have revealed
remarkably similar characteristics as a function of photon energy
loss: (1) the presence of a large, sharp and relatively
dispersionless peak centered around a few eVs, and (2) the
development of a low energy peak dispersive towards higher
frequencies for photon momentum
transfers from the Brillouin zone (BZ) center along either the BZ
edge or diagonal. Example of the data taken on
Ca$_{2}$CuO$_{2}$Cl$_{2}$\cite{Hasan2000} is shown in Figure 1.
The high energy peak has been associated with photon-induced
charge transfer between orbitals of different atoms\cite{Na} or
different orbitals of the same atom\cite{Hasan2000,Hasan2002},
while the low frequency peak has been associated with a transition
from the lower to upper Hubbard band across an effective Mott
gap\cite{Hasan2000,Hasan2002,Na} and a $q$-dependence of the Mott
gap has been inferred\cite{abbamonte}. However it does not seem
obvious why an excitation across a 
Mott gap would show dispersion given that the physics
of the Mott transition is highly local in character. 

Theoretical calculations on inelastic X-ray scattering
have to our knowledge been limited to energy-band model calculations
and exact diagonalization studies of
small clusters\cite{review-xray}. 
While energy-band calculations
might be appropriate for ground state properties of
weakly correlated systems they do not adequately address the role
of intra-atomic electron correlations which crucially affect
properties of the excitation spectra of strongly correlated systems.
Exact diagonalization studies of small clusters\cite{Hasan2000,1D}
largely focuses on the energy separation between the states
excitable by the X-rays such as excitons, and suffers the limitation
that the lineshape of the calculated spectra depend sensitively
on cluster size due to finite size effects on electron dynamics. Thus it
is clearly of interest for formulate a theory for inelastic X-ray
scattering which does not suffer from cluster-size effects and is
able to properly account for intra-atomic electron correlations.

\begin{figure}
\centerline{\psfig{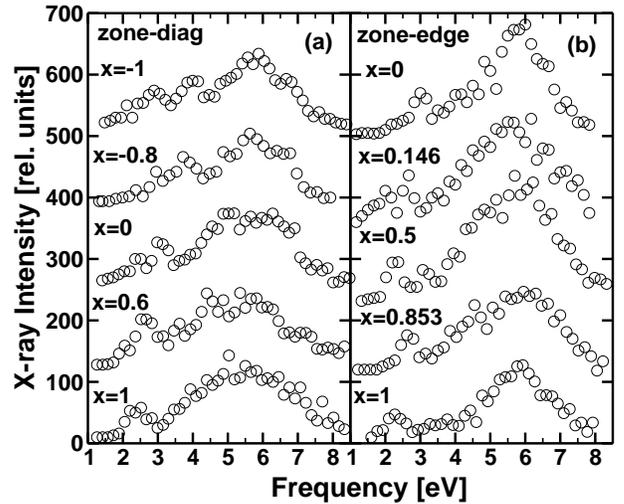}} \caption[] {
Experimental data for Ca$_{2}$CuO$_{2}$Cl$_{2}$ obtained in Ref.
{\protect\cite{Hasan2000}} for momentum transfers along the BZ
diagonal (a) and BZ edge (b), respectively. The values of 
the parameter $X=[\cos(q_{x}a)+\cos(q_{y}a)]/2$ indicate the
values of the momentum transfer $(q_{x},q_{y})$. Note that for panel
(b) the momentum transfer runs from $(0,0)$ to $(\pi,0)$ only, $X=1$ 
to $0$.
\label{fig: xray_experiment} }
\end{figure}

Two important features of the experimental data
have yet to be clarified. First,
the selection rules coming from the different orientations of the
{\it polarization} directions of the incoming and outgoing photons
as well as the direction of their scattered momenta have not been
used to determine the {\it symmetry} of the upper Hubbard band,
for example. These selection rules have led to intense
investigation of the dynamics of electrons in strongly correlated
systems like the high temperature superconductors to determine
information about charge dynamics on regions of the BZ or the
symmetry of the order parameter in the superconducting
state\cite{review}. Secondly, the mechanisms and/or pathways of
electronic relaxation revealed by inelastic X-ray scattering have
yet to be exploited. More precisely, besides the magnitude of the
matrix elements coupling the conduction band to the excited states
via the photon vector potential, the way in which the X-ray
induced charge
imbalance relaxes has remain largely unexplored. Inelastic X-ray
scattering affords an open window to investigate the symmetry and
pathway of charge dynamics in strongly correlated systems. This is
the topic of this paper.

We choose to focus attention on non-resonant X-ray
scattering, or scattering in which the frequency dependence of the
incoming or outgoing photons can be individually neglected and
only the frequency shift $\Omega=\omega_{i}-\omega_{s}$ enters,
where $\omega_{i,s}$ denotes incident, scattered X-ray energies, 
respectively. This certainly means that we have lost the ability
to make quantitative predictions concerning the overall
intensity of the
scattering and we cannot comment on lineshape changes induced by 
varying the incident photon frequency. However our goal is to
evaluate inelastic X-ray scattering in a
model in which the correlations can be handled exactly - the
Falicov-Kimball model in infinite dimensions - to determine
which features emerge from the strong correlations.

The Falicov-Kimball model, which has been used
to describe a variety of phenomenon in binary alloys\cite{FK},
contains itinerant band electrons and localized electrons, in
which the band electrons can hop with amplitude $t^{*}$ between
nearest neighbors and interact via a screened Coulomb interaction
$U$ with the localized electrons:
\begin{equation}
H=-{t^{*}\over{2\sqrt{d}}}\sum_{\langle i,j \rangle}c^{\dagger}_{i}c_{j}+E_{f}\sum_{i}w_{i}-\mu
\sum_{i}c^{\dagger}_{i}c_{i}+U\sum_{i}c^{\dagger}_{i}c_{i}w_{i},
\label{eq: one}
\end{equation}
where $c_{i}^{\dagger}, c_{i}$ is the spinless conduction electron
creation (annihilation) operator at site $i$ and $w_{i}=0$ or 1 is
a classical variable of the localized electron number at site $i$.
$E_{F}$ and $\mu$ control the filling of the localized and
conduction electrons, respectively. This is solved by dynamical
mean field theory as described in detail in Ref. \cite{brandt_mielsch}, where
the reader is referred to for details.

\begin{figure}
\centerline{\psfig{file=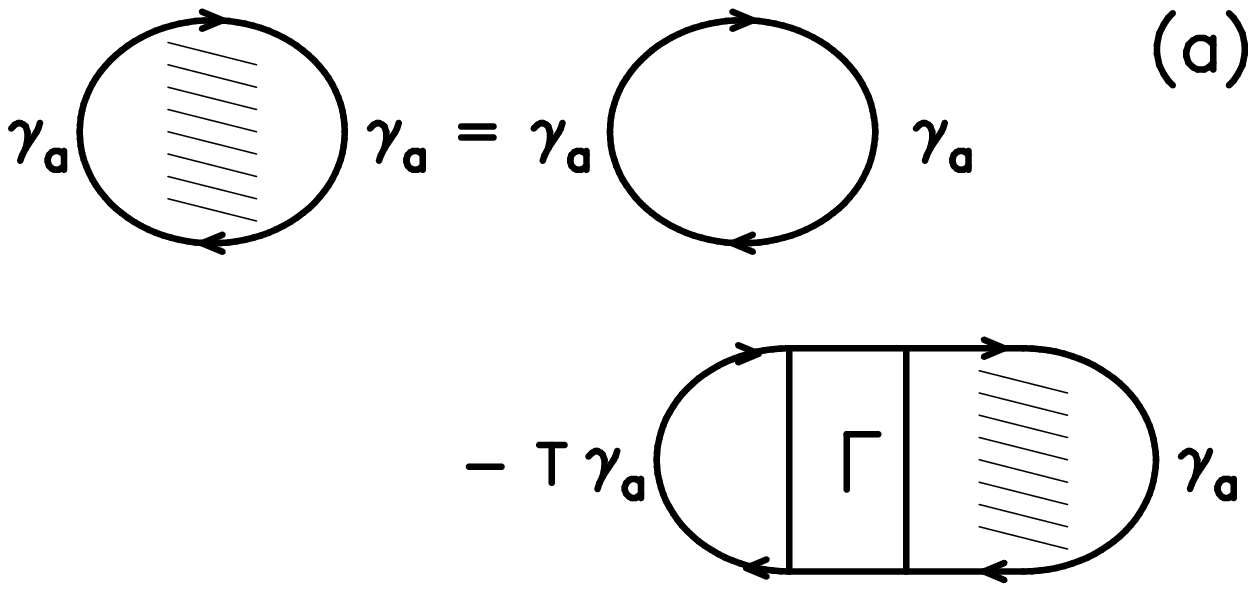,width=4cm,silent=}}
\centerline{\psfig{file=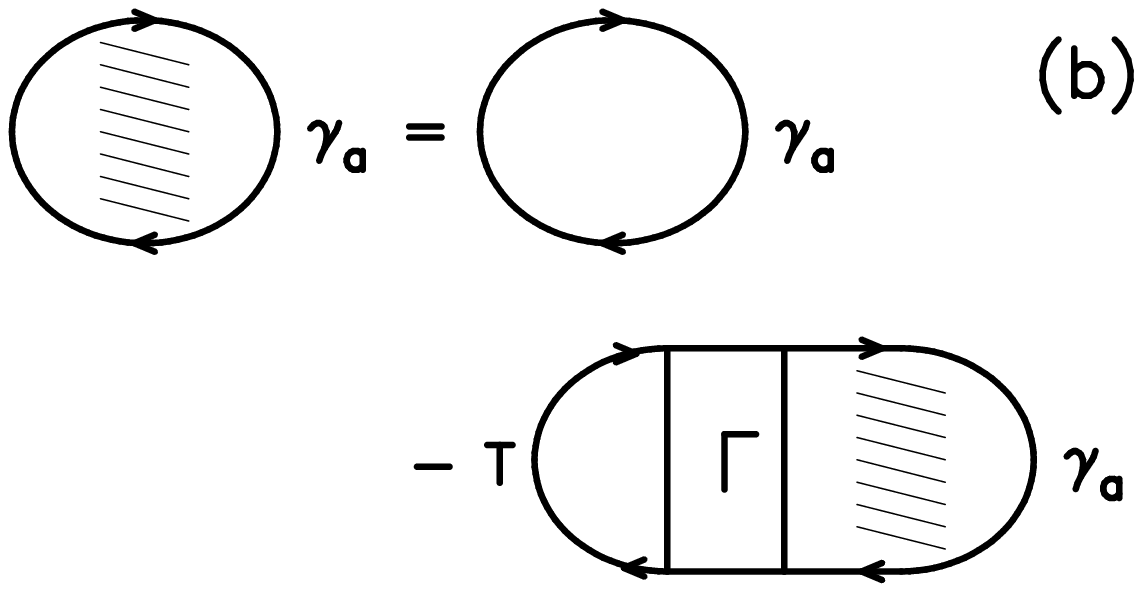,width=4cm,silent=}} \caption[]
{\label{fig: curr_curr_dyson} Coupled Dyson equations for the
inelastic X-ray scattering density-density correlation functions
described by the scattering amplitude $\gamma_a$.  Panel (a) depicts
the Dyson equation for the interacting correlation function, while
panel (b) is the supplemental equation needed to solve for the
correlation function. The symbol $\Gamma$ stands for the local
dynamical irreducible charge vertex given in Eq.~(\ref{eq:
dyn_vertex_final}). In situations
where there are no charge vertex corrections, the correlation
function is simply given by the first (bare-bubble) diagram on the
right hand side of panel (a). }
\end{figure}

In this single band model with energy $\epsilon({\bf k})$, the inelastic
X-ray response is given formally by an effective density-density
correlation function $S({\bf q},\omega)=-{1\over{\pi}}[1+n(\omega)]
\chi^{\prime\prime}({\bf q},\omega)$ with
\begin{equation}
\chi({\bf q},\omega)=\langle [\tilde\rho({\bf q}),\tilde\rho(-{\bf q})]
\rangle_{(\omega)}
\end{equation}
formed with an ``effective'' density operator given by
\begin{equation}
\tilde\rho({\bf q})=\sum_{\bf k,\sigma}\gamma_{a}({\bf k})
c_{\sigma}^{\dagger}({\bf k+q/2})c_{\sigma}({\bf k-q/2}).
\end {equation}
The strength of the scattering $\gamma_{a}$ is determined by the
curvature of the band as
\begin{equation}
\gamma_{a}({\bf k})=\sum_{\alpha,\beta}e_{\alpha}^{s}
{\partial^{2}\epsilon({\bf k})\over{\partial k_{\alpha} \partial k_{\beta}}}
e_{\beta}^{i}.
\end{equation}
Here ${\bf e^{i,s}}$ denote the incident, scattered X-ray polarization
vectors, respectively, and 
we have chosen units $k_{B}=c=\hbar=1$ and have set the hypercubic
lattice constant equal to 1.
We can classify the scattering amplitudes by point group
symmetry operations. If we choose $e^i=(1,1,1,...)$ and
$e^s=(1,-1,1,-1,...)$, then we have the $B_{\rm 1g}$ sector, while
$e^{i}=e^{s}=(1,1,1,...)$ projects out the $A_{1g}$ sector
since the $B_{2g}$ component is identically zero in our model due
to the inclusion of only nearest-neighbor hopping.
We thus can cast the scattering amplitudes in a simple form:
$\gamma_{A_{\rm 1g}}
(\textbf{k})=-\epsilon(\textbf{k})$ and $\gamma_{B_{\rm
1g}}(\textbf{k})= t^*\sum_{j=1}^\infty \cos \textbf{k}_j
(-1)^j/\sqrt{d}$. 

The Dyson equation for the density-density correlation function
takes the form given in Fig.~\ref{fig: curr_curr_dyson}. Note that
there are two coupled equations illustrated in Figs.~\ref{fig:
curr_curr_dyson}~(a) and (b); these equations differ by the number
of $\gamma_a$ factors in them. The irreducible vertex function
$\Gamma$ is the dynamical charge vertex~\cite{charge_vertex} which
takes the form
\begin{equation}
\Gamma(i\omega_m,i\omega_n;i\nu_{l\ne 0})=
\delta_{mn}\frac{1}{T}\frac{\Sigma_m-\Sigma_{m+l}}{G_m-G_{m+l}}.
\label{eq: dyn_vertex_final}
\end{equation}
on the imaginary axis
[$i\omega_m=i\pi T(2m+1)$ is the Fermionic Matsubara frequency and
$i\nu_l=2i\pi T l$ is the Bosonic Matsubara frequency]. Here
$\Sigma_m=\Sigma(i\omega_m)$ is the local self energy on the
imaginary axis and $G_m=G(i\omega_m)$ is the local Green's
function on the imaginary axis. If the scattering amplitude
$\gamma$ does not have a projection onto the full
symmetry of the lattice, then there are no vertex corrections from
the local dynamical charge vertex~\cite{khurana}.

\begin{figure}
\centerline{\psfig{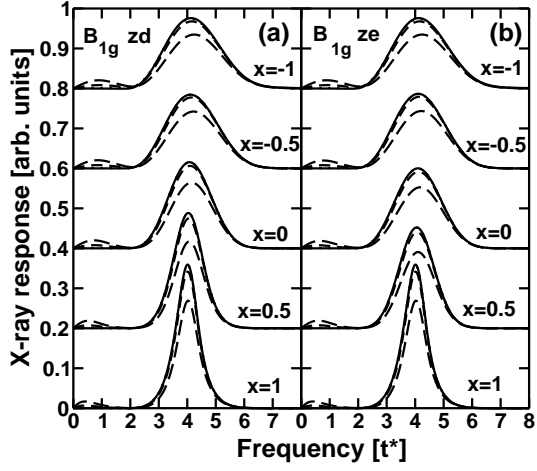}}
\caption[]{\label{fig: xray_fk} Inelastic X-ray scattering
response in the $B_{\rm 1g}$ channel along (a) the Brillouin zone
diagonal and (b) along the zone edge for the half-filled
Falicov-Kimball model on a hypercubic lattice. 
The solid, dotted,
short-dashed and long-dashed curves
correspond to temperatures $T=1.0,$ 0.5, 0.25, 0.1, respectively. }
\end{figure}

A straightforward calculation, shows that the $B_{\rm 1g}$
response has no vertex corrections on the zone diagonal
$\textbf{q}=(q,q,q,q,...)$.  Hence, the $B_{\rm 1g}$ response is
the bare bubble: 
\begin{eqnarray}
\chi_{B_{\rm 1g}}(\textbf{q},\nu)&=&\frac{i}{4\pi}\int_{-\infty}^{\infty}
d\omega\bigl \{ f(\omega)\chi_0(\omega;X,\nu)-f(\omega+\nu)
\nonumber\\
&\times&\chi_0^*(\omega;X,\nu)
[f(\omega)-f(\omega+\nu)]\tilde\chi_0(\omega;X,\nu) \bigr \}\nonumber\\
\label{eq: b1g_final}
\end{eqnarray}
with
\begin{eqnarray}
\chi_0(\omega;X,\nu)&=&-\int_{-\infty}^{\infty}d\epsilon\rho(\epsilon)
\frac{1}{\omega+\mu-\Sigma(\omega)-\epsilon}\frac{1}{\sqrt{1-X^2}}\cr
&\times&F_\infty \left (
\frac{\omega+\nu+\mu-\Sigma(\omega+\nu)-X\epsilon}{\sqrt{1-X^2}}
\right ) , \label{eq: chi0}
\end{eqnarray}
and
\begin{eqnarray}
\tilde\chi_0(\omega;X,\nu)&=&-\int_{-\infty}^{\infty}d\epsilon\rho(\epsilon)
\frac{1}{\omega+\mu-\Sigma^*(\omega)-\epsilon}\frac{1}{\sqrt{1-X^2}}\cr
&\times&F_\infty \left (
\frac{\omega+\nu+\mu-\Sigma(\omega+\nu)-X\epsilon}{\sqrt{1-X^2}}
\right ) . \label{eq: chi0tilde}
\end{eqnarray}
Here we have used the following notation: $f(\omega)=1/[1+\exp
(\omega)]$ is the Fermi-Dirac distribution,
$\rho(\epsilon)=\exp(-\epsilon^2)/ \sqrt{\pi}$ is the
noninteracting density of states; $\Sigma(\omega)$ is the local
self energy on the real axis; $X=\lim_{d\rightarrow\infty}
\sum_{i}\cos q_{i}/d=\cos(q)$  for the zone-diagonal
wavevector $\textbf{q}=(q,q,q,...,q)$; and $F_\infty(z)= \int
d\epsilon \rho(\epsilon)/(z-\epsilon)$ is the Hilbert transform of
the noninteracting density of states.  Techniques for finding the
self energy ~\cite{brandt_mielsch} have appeared elsewhere.

The $A_{\rm 1g}$ response
everywhere and the $B_{\rm 1g}$ response off of the zone diagonal,
do have vertex corrections.  The calculation of each response
function is straightforward, but tedious.  One needs to first
solve the coupled equations depicted in Fig.~\ref{fig:
curr_curr_dyson} on the imaginary axis and then perform the
analytic continuation as in the Raman scattering
case~\cite{raman_long}. The end result is cumbersome and will be
presented in a longer paper.

\begin{figure}
\centerline{\psfig{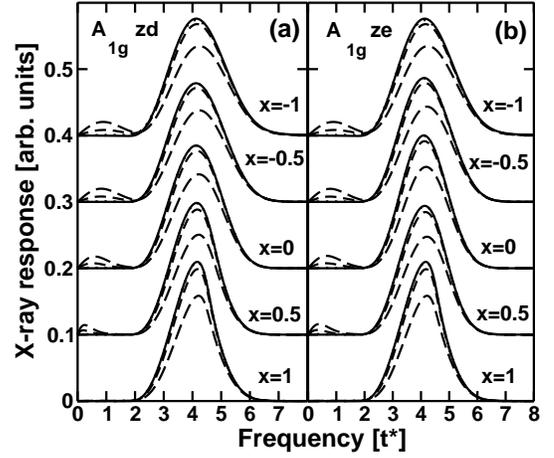}}
\caption[]{\label{fig: xray_fk2} Inelastic X-ray scattering
response in the $A_{\rm 1g}$ channel along (a) the Brillouin zone
diagonal and (b) along the zone edge for the half-filled
Falicov-Kimball model on a hypercubic lattice. The solid, dotted,
short-dashed and long-dashed curves
correspond to temperatures $T=1.0, 0.5, 0.25, 0.1,$ respectively. }
\end{figure}

The results for a correlated insulator $U=4t^{*}$ at different
temperatures are shown in Figs. ~\ref{fig: xray_fk} 
and  ~\ref{fig: xray_fk2} 
for $B_{1g}$ and $A_{1g}$ inelastic X-ray scattering,
respectively, as a function of transfered energy for different momentum
transfers throughout the BZ measured by
the factor $X$. Panel (a) for Figs. ~\ref{fig: xray_fk} 
and  ~\ref{fig: xray_fk2} refer to scattering along
the zone diagonal $X=\cos q$ for the zone-diagonal
wavevector $\textbf{q}=(q,q,q,...,q)$, and panel (b) refer to
scattering along the zone edge
[here we have $\textbf{q} =(q,0,q,0,...,q,0)$ for $1\ge X=(1+\cos
q)/2\ge 0$ and $\textbf{q} =(\pi,q,\pi,q,...,\pi,q)$ for $0\ge
X=(-1+\cos q)/2\ge -1$ 
The curves have been shifted vertically for clarity.
The lowest set of curves $X=1$ correspond to Raman scattering with
optical photons~\cite{raman_long}. The main qualitative 
feature in both Figures are the presence
of a small, dispersive low-energy peak for frequencies $\sim t^{*}$
and a large, dispersionless charge-transfer peak $\sim U$.
While the charge-transfer peak remains relatively robust with 
increasing temperature, the low energy peak gains intensity from zero
as temperature is increased. In particular all momenta show the development of
low-energy spectral weight as $T$ increases and there is a non-dispersive 
isosbestic point - a frequency at which the spectra are temperature
independent - around $\sim U/2$. The high energy peak reflects the energy
scale for excitations across the Mott gap and is relatively dispersionless
due to the local nature of the correlations. In contrast
the low energy feature is a consequence 
thermally generated double occupancies which open up a low energy
band up to energies $~\sim t^{*}$ able to scatter X-rays. 
For decreasing temperature the low energy intensity 
disappears and only scattering across the Mott gap remains.

The charge-transfer peak is sharp
near the BZ center ($X=1$), but
broadens for momentum transfers approaching the BZ corner $X=-1$, more
so for the $B_{1g}$ channel than for $A_{1g}$.
In fact $A_{1g}$ and $B_{1g}$ are 
identical at the $(\pi,\pi,...,\pi)$ point $X=-1$ due to the  
local approximation. Any variation in the signal at the
zone corner in different symmetry channels must be due to nonlocal
many-body correlations.  

An important difference is that the $A_{\rm 1g}$ results
have no low-energy spectral weight for $\textbf{q}=0$, corresponding
to inelastic Raman scattering\cite{raman_long}. The vertex
corrections remove all remnants of the low-energy response here,
but it enters for any finite value of \textbf{q}. For an unpolarized
measurement, the X-ray response is a superposition of the
$B_{1g}$ and $A_{1g}$ spectra. 

\begin{figure}
\centerline{\psfig{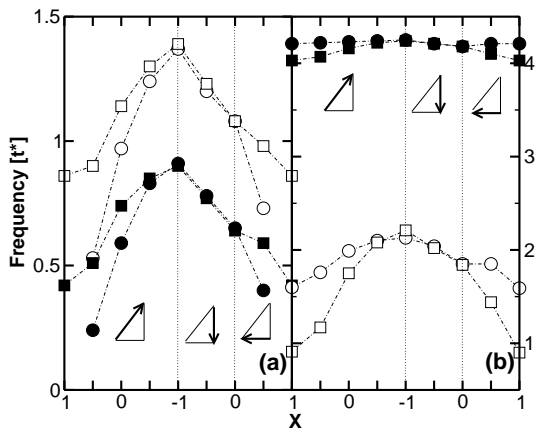}}
\caption[]{\label{fig: results} Plots of the low energy (panel a)
and charge transfer (panel b) peak
positions (solid symbols) and 
broadening (full width at half maximum, open symbols) for $T=t^{*}$
determined from Figs. 3 and 4 for $A_{1g}, B_{1g}$ (circles, squares),
respectively.
}
\end{figure}

We plot the behavior of the peak position and peak width (full-width
at half maximum) for both the low-energy peak and the charge-transfer
peak for both channels in Fig. ~\ref{fig: results}. One can see more
clearly that the low-energy peak has a width larger than its energy
for both channels and for the $A_{1g}$ channel follows
the behavior of the corresponding $B_{1g}$ feature
away from the zone center.
The charge transfer peak on the other hand is well defined for both
channels. The only dispersive feature of the charge-transfer peak
is the width of the $B_{1g}$ peak.

Referring back to the experimental data shown in Fig. 
~\ref{fig: xray_experiment}, it is tempting to associate the 
relatively dispersionless high
energy peak with an excitation across a
charge-transfer gap and the broad
low energy peak with the dispersive feature generated from double 
occupancies. However the experimental data is not yet complete
as the polarization
and temperature dependence have not been measured. Our
theory would predict, if this interpretation were correct, that the
low energy feature would decrease in intensity as temperature is
lowered. Moreover a polarization-dependent measurement could perhaps
deconvolve the high-energy peak into two separate peaks of $A_{1g}$ and
$B_{1g}$ symmetry, and would also be able to separate different
behavior of the low energy peak near the zone center.
Thus
we believe it would
be quite interesting to examine inelastic X-ray scattering at
well-controlled temperatures and with polarizers for the incident
and scattered light.  We believe that a number of new and
interesting features of charge excitations in correlated systems
are likely to emerge if this can be accomplished.

We would like to thank M. Z. Hasan and Z.-X. Shen for
sharing their experimental data with us and thank them
for valuable discussions.  J.K.F.
acknowledges support from the NSF under grant number DMR-9973225.
T.P.D. acknowledges support by NSERC.

\end{document}